\documentclass[12pt]{iopart}
\usepackage{epsfig,wrapfig}

\newcommand{\beq}{\begin{equation}}
\newcommand{\eeq}{\end{equation}}
\newcommand{\lsm}{\stackrel{\scriptstyle <}{\phantom{}_{\sim}}}

\newcommand{\tri}{\mbox{\scriptsize tr}}

\newcommand{\lb}{\mbox{\scriptsize LB}}

\begin{document}

\title[Mechanisms driving alteration of the Landau state \dots]%
{Mechanisms driving alteration of the Landau state
in the vicinity of a second-order phase transition}

\author{M Baldo\dag, V V Borisov\ddag, J W Clark\S, V A Khodel\ddag \\
and M V Zverev\ddag}


\address{\dag\
Istituto Nazionale di Fisica Nucleare, Sezione di Catania,
I-95123, Catania, Italy }

\address{\ddag\
Russian Research Center Kurchatov Institute,
Moscow, 123182 Russia }

\address{\S\
McDonnell Center
for the Space Science and Department of Physics,
Washington University, St.Louis, MO 63130, USA  }

\begin{abstract}
The rearrangement of the Fermi surface of a homogeneous Fermi
system upon approach to a second-order phase transition is
studied at zero temperature.  The analysis begins with an
investigation of solutions of the equation $\epsilon(p)=\mu$,
a condition that ordinarily has the Fermi momentum $p_F$ as
a single root.  The emergence of a bifurcation point in this
equation is found to trigger a qualitative alteration of the Landau
state, well before the collapse of the collective degree of freedom
that is responsible for the second-order transition.  The
competition between mechanisms that drive rearrangement of the
Landau quasiparticle distribution is explored, taking into account
the feedback of the rearrangement on the spectrum of critical
fluctuations.  It is demonstrated that the transformation of
the Landau state to a new ground state may be viewed as a
first-order phase transition.
\end{abstract}

\date{\today}



\ead{zverev@mbslab.kiae.ru}
\maketitle

\section{Introduction}

In the Landau-Migdal theory of Fermi liquids \cite{lan,mig0},
the ground state of a homogeneous Fermi system is described
in terms of a quasiparticle momentum distribution $n_F(p,T)$
that coincides with the momentum distribution of the ideal
Fermi gas.  This theory has been remarkably successful in
advancing our qualitative and quantitative understanding
of a broad spectrum of Fermi systems, including bulk liquid
$^3$He, conventional superconductors, and nucleonic subsystems
in neutron stars.  However, the theory is known to fail in
the strongly correlated electron systems present in high-$T_c$
compounds.  Certain experimental results obtained very recently
\cite{cas,krav,pud} may prove decisive to an understanding of
this failure.  The systems involved are a dilute two-dimensional
(2D) electron gas and 2D liquid $^3$He.  The experiments show how,
under variation of the density, these systems progress from conditions
of moderate correlation to the regime of very strong correlation.
A striking feature is that both systems appear to experience a
divergence of the effective mass $M^*$ as the density approaches
a critical value $\rho_{\infty}$ associated with some kind of
phase transition, which is presumably of second order \cite{pud}.

We base our analysis on a necessary condition for the stability
of the Landau state, namely that the change $\delta E_0$ of the ground
state energy $E_0$ remain positive for any admissible variation
from the quasiparticle distribution $n_F(p)=\theta(p_F-p)$,
while keeping the particle number unchanged.  Explicitly, this
condition reads
\beq
\delta E_0= \int \xi(p;n_F)\delta n_F({\bf p})d\tau > 0\ ,
\label{nsc}
\eeq
for any variation $\delta n_F({\bf p})$ satisfying
\beq
\int \delta n_F({\bf p})d\tau =0 \ .
\eeq
In these equations, $d\tau$ is the volume element in momentum space,
while $\xi(p)=\epsilon(p)-\mu$ is the single-particle (sp) spectrum,
measured from the chemical potential $\mu$ and evaluated with the
distribution $n_F(p)$.

The condition (\ref{nsc}) holds provided the equation
\beq
\xi(p)=0
\label{root}
\eeq
has the single root $p=p_F$.  Otherwise, it is violated, the
Landau state loses its stability, and the ground state must
take another form, implying a rearrangement of single-particle
degrees of freedom.  In weakly correlated Fermi systems, $\xi(p)$
is a monotonic function of $p$, so that equation~(\ref{root}) has no
extra roots.  However, as correlations build up, the character of
the curve $\xi(p)$ may change.  Indeed, it becomes non-monotonic in
the vicinity of an impending second-order phase transition, when
critical fluctuations of wave number $q_c>0$ produce a diverging
susceptibility and hence a collapse of the corresponding
collective degree of freedom.

\begin{figure}
\begin{center}
\mbox{\epsfig{file=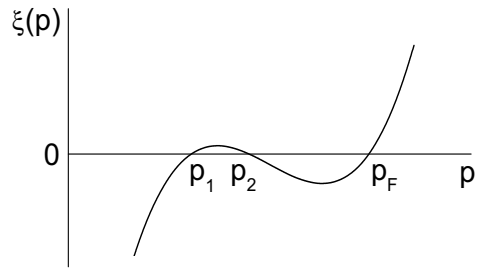,width=0.4\linewidth,height=0.25\linewidth}}
\end{center}
\caption{Illustration of the emergence of additional
roots $p_1,\,p_2$ of equation~(\ref{root}).
}
\label{fig:max}
\end{figure}

Let the second-order phase transition occur at a critical density $\rho_c$.
As we shall see, there is another critical density $\rho_b$ at which
a bifurcation arises in equation~(\ref{root}), resulting in the emergence
of a two additional roots $p_1,p_2$ (see figure~\ref{fig:max}).  The
distance between these extra roots increases linearly from zero in
proportion to $|\rho-\rho_b|$.  It should be emphasized that the stability
condition (\ref{nsc}) is never violated when applied to variations
of the quasiparticle distribution $n(p)$ for momenta lying beyond
the interval $[p_1,p_2]$.  Hence, at $|\rho-\rho_b|\ll\rho$ the
rearrangement process is confined to a constricted domain in momentum
space.  Accordingly, a rearrangement that entails a major alteration
of the ground state in configuration space, involving all of the occupied
sp states and therefore disfavored energetically and therefore irrelevant
to the present study.  In particular, Mott-Hubbard localization
is ruled out.

For this reason our attention will be focused on two plausible
scenarios for the rearrangement of the momentum distribution $n_F(p)$.
In the first scenario, modification of the Landau state consists in
the formation of empty spaces in momentum space that have been named
Lifshitz bubbles (LB).  In the LB phase, the quasiparticle occupation
numbers have the usual values 0 and 1, but the Fermi surface becomes
multi-connected.  In fact, this and related phenomena were studied in
model problems more than 20 years ago \cite{vary,aguil}.  In the limit
$|\rho-\rho_b|\ll\rho$, the LB mechanism has no rivals, provided the
interval $[p_1,p_2]$ is not located in the immediate vicinity of the
Fermi momentum $p_F$.  Otherwise, there exists a novel competitor
called fermion condensation \cite{ks,vol,noz,physrep,ksz}, which
is the second scenario to be examined here.  Fermion condensation
is a rearrangement of the Landau state leading from the Fermi step
$n_F(p)$ to a {\it continuous} quasiparticle momentum distribution
$n(p)$ having no Migdal jump at $p_F$.  In the region C adjacent to
the original Fermi surface where $n(p)$ departs from $n_F(p)$ by dropping
smoothly from 1 to 0, the sp spectrum turns out to be completely
flat, with $\epsilon(p) = \mu$.  This behavior gives rise to
a singular, $\delta$-function term in the density of states
$\rho(\varepsilon)$.  Considered as a phase transition, fermion
condensation does not break any symmetry, and has much in common
with the classical gas-liquid phase transition \cite{physrep}.
However, the presence of the singularity in $\rho(\varepsilon)$
enhances the feedback of the rearrangement process on the spectrum
of the relevant critical fluctuations, which, in its turn, affects
the competition between the two mechanisms proposed for rearrangement
of the Landau state.

After investigating the nature of the instability of the Landau
state, we shall illuminate the competition between LB and FC
rearrangement scenarios by considering a simple model, in which the
softening effect is assumed to depend linearly on the phase volume
of region C occupied by the fermion condensate.  It will be found
that formation of the FC state exerts the greater impact on the
collective degree of freedom.  This being the case, we demonstrate that
(i) the FC phase wins the contest with the LB reconfiguration, and
(ii) the corresponding transformation of the Landau state is a
first-order phase transition.

\section{Instability of the Landau state }

To gain detailed insight into the emergence of the bifurcation point
in equation~(\ref{root}), we employ the Landau relation \cite{lan,trio}
\beq
{\partial\xi(p)\over\partial {\bf p}} =
  {{\bf p}\over M} +
     \int f({\bf p},{\bf p}_1)
         {\partial n(p_1)\over\partial {\bf p}_1}d\tau_1 \ ,
\label{pgr}
\eeq
which connects the quasiparticle group velocity $d\xi/dp$ with the
momentum distribution $n(p)$ in terms of the Landau scattering
amplitude $f$.

First, we consider the case $q_c\sim p_F$, which applies to several
phase transitions of fundamental interest.  One of these is pion
condensation, predicted to occur in (3D) neutron matter due to
collapse of the collective spin-isospin mode with pion quantum
numbers \cite{mig1,mig2,pand,wir}.  In this situation, the leading
term in the amplitude $f$, being proportional to the singular term
in the static spin-isospin susceptibility, has the form \cite{mig1}
\beq
f(q)= {g \over \kappa^2(\rho)+(q^2/q^2_c-1)^2} \ ,
\label{tio}
\eeq
where $g$ is a positive coupling constant and the stiffness coefficient
$\kappa^2(\rho)$ vanishes at the critical density $\rho_c$.  The same
form of $f$ is expected to apply in two-dimensional liquid $^3$He,
where spin fluctuations play an important role \cite{krot}.

The sp spectrum $\xi(p)$ in the Landau state, with quasiparticle
distribution $n_F(p)$, may be evaluated in closed form by means
of equation~(\ref{pgr}).  Substituting the expression (\ref{tio})
for the amplitude $f$ and performing the integration on the
right-hand side, we obtain
\begin{eqnarray}
\fl
{d\xi(p,n_F)\over dp}={p\over M}+{gq^4_c\over 16\pi^2 p^2}
\Biggl[ {1\over 2} \log
\frac
  {[(p{-}p_F)^2{-}q^2_c]^2+\kappa^2 q_c^4}
  {[(p{+}p_F)^2{-}q^2_c]^2+\kappa^2 q_c^4}
\nonumber \\
\lo + \frac{p^2{+}p_F^2{-}q_c^2}{\kappa q_c^2}
\left(
  \arctan\frac{(p_F{+}p)^2{-}q^2_c}{\kappa q^2_c}
 -\arctan\frac{(p_F{-}p)^2-q^2_c}{\kappa q^2_c}
\right)
\Biggr] \ .
\label{deriv}
\end{eqnarray}
Further integration yields the formula
\beq
\xi(p)=\xi^0(p) + {gq_c^2p_F\over 8\pi^2\kappa}\,w(p) \ ,
\label{specpion}
\eeq
the dimensionless function $w(p)$ being given by
\begin{eqnarray}
\fl
w(p) =
\frac{p^2_F{-}q^2_c{-}p^2}{2pp_F}\,
\Biggl[
\arctan\frac{(p_F{+}p)^2{-}q^2_c}{\kappa q^2_c}
-\arctan\frac{(p_F{-}p)^2-q^2_c}{\kappa q^2_c}
\Biggr]
\nonumber \\
\fl \qquad
- \frac{\kappa q^2_c}{4pp_F}
\ln\frac {[(p_F{+}p)^2{-}q^2_c]^2{+}\kappa^2 q^4_c}
         {[(p_F{-}p)^2{-}q^2_c]^2{+}\kappa^2 q^4_c}
\nonumber \\
\fl \qquad
+ \frac{\kappa q^2_c}{4\sigma_{+}p_F}
\Biggl[
\ln\left|\frac
       {(p_F{+}p)^2{-}2\sigma_{+}(p_F{+}p){+}\sigma^2_{0}}
       {(p_F{+}p)^2{+}2\sigma_{+}(p_F{+}p){+}\sigma^2_{0}}
  \right|
+\ln\left|\frac
       {(p_F{-}p)^2{-}2\sigma_{+}(p_F{-}p){+}\sigma^2_{0}}
       {(p_F{-}p)^2{+}2\sigma_{+}(p_F{-}p){+}\sigma^2_{0}}
  \right|
             \Biggr]
\nonumber \\
\fl \qquad
+ \frac{\kappa q^2_c}{2\sigma_{-}p_F}
\Biggl[
    \arctan\frac{p_F{+}p{+}\sigma_{+}}{\sigma_{-}}
+   \arctan\frac{p_F{-}p{+}\sigma_{+}}{\sigma_{-}}
\nonumber \\
\qquad\qquad
+ \arctan\frac{p_F{+}p{-}\sigma_{+}}{\sigma_{-}} +
    \arctan\frac{p_F{-}p{-}\sigma_{+}}{\sigma_{-}})
\Biggr] \ ,
\label{www}
\end{eqnarray}
where
\beq
\sigma_{\pm}= q_c\,
   \left(\frac{\sqrt{1{+}\kappa^2}{\pm}1}{2} \right)^{1/2} ,
        \qquad    \qquad
      \sigma_0=q_c(1+\kappa^2)^{1/4} \ ,
\label{sigma}
\eeq
and $\xi^0(p)=p^2/2M-\mu$.

\begin{figure}
\begin{center}
\mbox{\epsfig{file=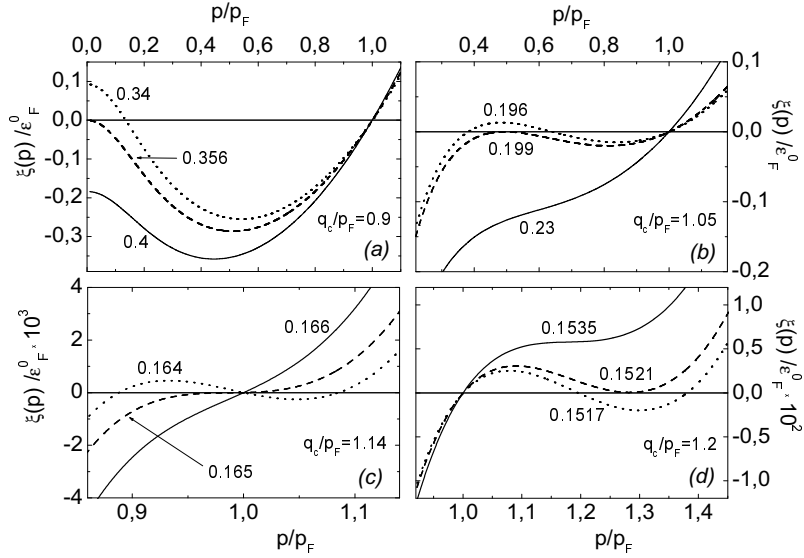,width=0.7\linewidth,height=0.5\linewidth}}
\end{center}
\caption{
Neutron quasiparticle spectra $\xi(p)$ (in units of
$\varepsilon^0_F$) evaluated for $q_c=0.9\, p_F$ (panel (a)),
$q_c=1.05\, p_F$ (panel (b)), $q_c=1.14\, p_F$ (panel (c)), and
$q_c=1.2\, p_F$ (panel (d)).  Corresponding values of the
parameter $\kappa$ are indicated near the curves.
}
\label{fig:nmxi}
\end{figure}

\begin{figure}
\begin{center}
\mbox{\epsfig{file=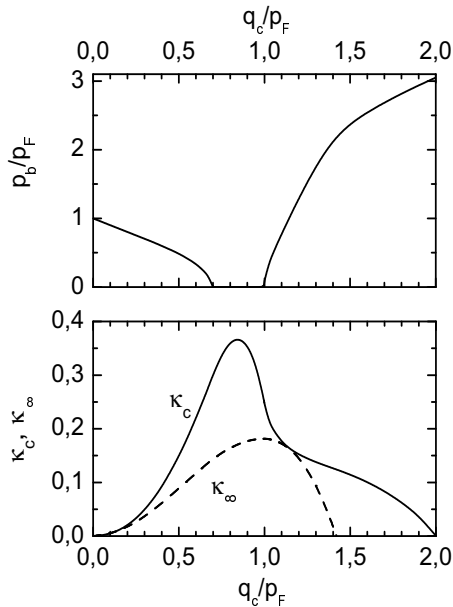,width=0.39\linewidth,height=0.52\linewidth}}
\end{center}
\caption{
Upper panel: Position of the bifurcation point $p_b$ in units of the
Fermi momentum, versus the critical wave number $q_c$ (also in units
of $p_F$).  Lower panel: critical parameters $\kappa_b$ and $\kappa_{\infty}$
as functions of $q_c/p_F$.
}
\label{fig:nmbe}
\end{figure}

Results of numerical calculations for neutron matter are shown
in figures~\ref{fig:nmxi} and \ref{fig:nmbe}.  For simplicity,
we take the coupling constant in the amplitude (\ref{tio})
to be $g=1/2m^2_{\pi}$, corresponding to bare $\pi NN$ vertices.
The spectrum $\xi(p)$, evaluated with the critical momentum
$q_c=0.9\,p_F$ and for four values of the parameter $\kappa$,
is displayed in panel (a) of figure~\ref{fig:nmxi}.

A new root $p_b\sim 0$ of equation~(\ref{root}) is seen to appear
at $\kappa_b\simeq 0.356$, signaling that the Fermi step has become
unstable.  It is worth noting that at the customary values \cite{mig1}
of the critical momentum, $q_c/p_F \sim 0.7-1.0$, the bifurcation
point lies exactly at the origin in $p$.  However, as $q_c$ increases
to greater values, it rapidly moves toward the Fermi momentum and
leaves the Fermi sphere at $q_c\sim 1.14\,p_F$.  This evolution
is illustrated by panels (b)--(d) of figure~\ref{fig:nmxi},
where the spectra $\xi(p)$ calculated for $q_c=1.05\,p_F, 1.14\,p_F$,
and $1.2\,p_F$ are drawn.  Figure~\ref{fig:nmbe} depicts the
dependence $p_b(q_c)$ in the large interval $0<q_c<2\,p_F$ (upper
panel), together with the dependence of the critical parameter
$\kappa_b$ on the wave number $q_c$ (lower panel).  Remarkably, the
largest values of $\kappa_b$ are achieved just in the preferred range
$q_c/p_F\sim 0.7-1.0$. The value $\kappa_{\infty}$ of $\kappa$
at which the border of the instability region $[p_1,p_2]$ reaches
the Fermi momentum $p_F$, is plotted in the lower panel of
figure~\ref{fig:nmbe}.  The resulting curve lies below the
curve of $\kappa_b(q_c)$ everywhere except for the point of
contact at $q_c\simeq 1.14\,p_F$.

\begin{figure}
\begin{center}
\mbox{\epsfig{file=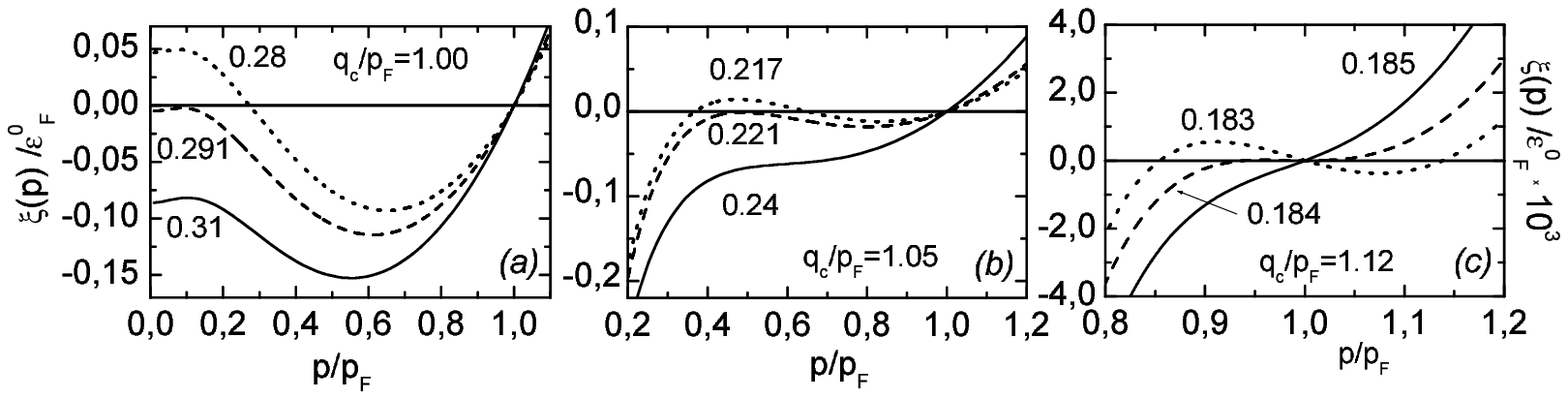,width=1.\linewidth,height=0.31\linewidth}}
\end{center}
\caption{
Quasiparticle spectrum $\xi(p)$ of 2D liquid $^3$He $\xi(p)$
(in units of $\varepsilon^0_F$), calculated for $q_c=1.0 p_F$ (panel
(a)), $q_c=1.05 p_F$ (panel (b)), and $q_c=1.12 p_F$ (panel (c)).
The values of $\kappa$ are indicated near the curves.
}
\label{fig:2dhexi}
\end{figure}

The above results refer to the 3D problem.  In the 2D case, analytical
evaluation of the spectrum $\xi(p)$ is rather cumbersome, but its
numerical computation is easily accomplished.  We have calculated
$\xi(p)$ for 2D liquid $^3$He under the assumption that the dominant
term in the quasiparticle amplitude is by governed by the static
spin-spin susceptibility.  Results are shown in figure~\ref{fig:2dhexi}.
While the spectrum of 2D liquid $^3$He is found to differ quantitatively
from that of 3D neutron matter, the shapes are qualitatively similar,
as is the evolution with increasing $q_c$.

We infer from these two sets of results that in the case
$q_c\lsm p_F$, the Landau state becomes unstable prior to the
second-order phase transition itself.  As will be seen, this
is a generic feature.  On the other hand, particulars of the
alteration of the Landau state will depend on the parameters
that specify the amplitude $f$. To illustrate the general
situation, we focus on a phase transition associated with
the spontaneous generation of density waves in dense neutron
matter or the dilute electron gas, in both of which the critical
wave number $q_c$ is close to $2p_F$.  In this case, the scattering
amplitude $f$ has the same form as (\ref{tio}), but the sign of
$g$ is negative \cite{physrep}.

\begin{figure}
\begin{center}
\mbox{\epsfig{file=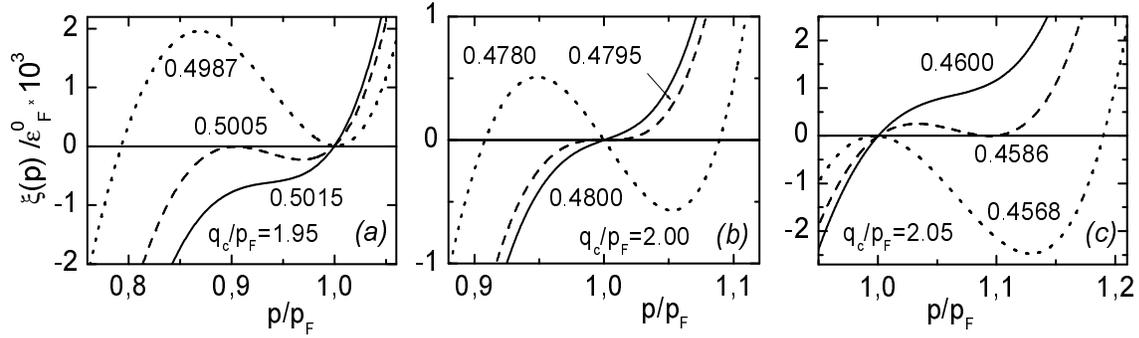,width=1.\linewidth,height=0.31\linewidth}}
\end{center}
\caption{
Electron spectra $\xi(p)$ in 3D (measured in units of
$\varepsilon^0_F$), as calculated for $q_c=1.95\,p_F$
(panel (a)), $q_c=2.00\,p_F$ (panel (b)), and $q_c=2.05\,p_F$
(panel (c)). The corresponding values of the parameter
$\kappa$ are indicated near the curves. In each panel,
the solid line traces the spectrum before the instability point
is attained ($\kappa>\kappa_b$), and the dotted line shows that
at $\kappa<\kappa_b$. In the panels (a) and (c),
the dotted line indicates the spectrum for $\kappa=\kappa_{\infty}$,
at which the instability region reaches the Fermi surface.
}
\label{fig:3degxi}
\end{figure}

\begin{figure}
\begin{center}
\mbox{\epsfig{file=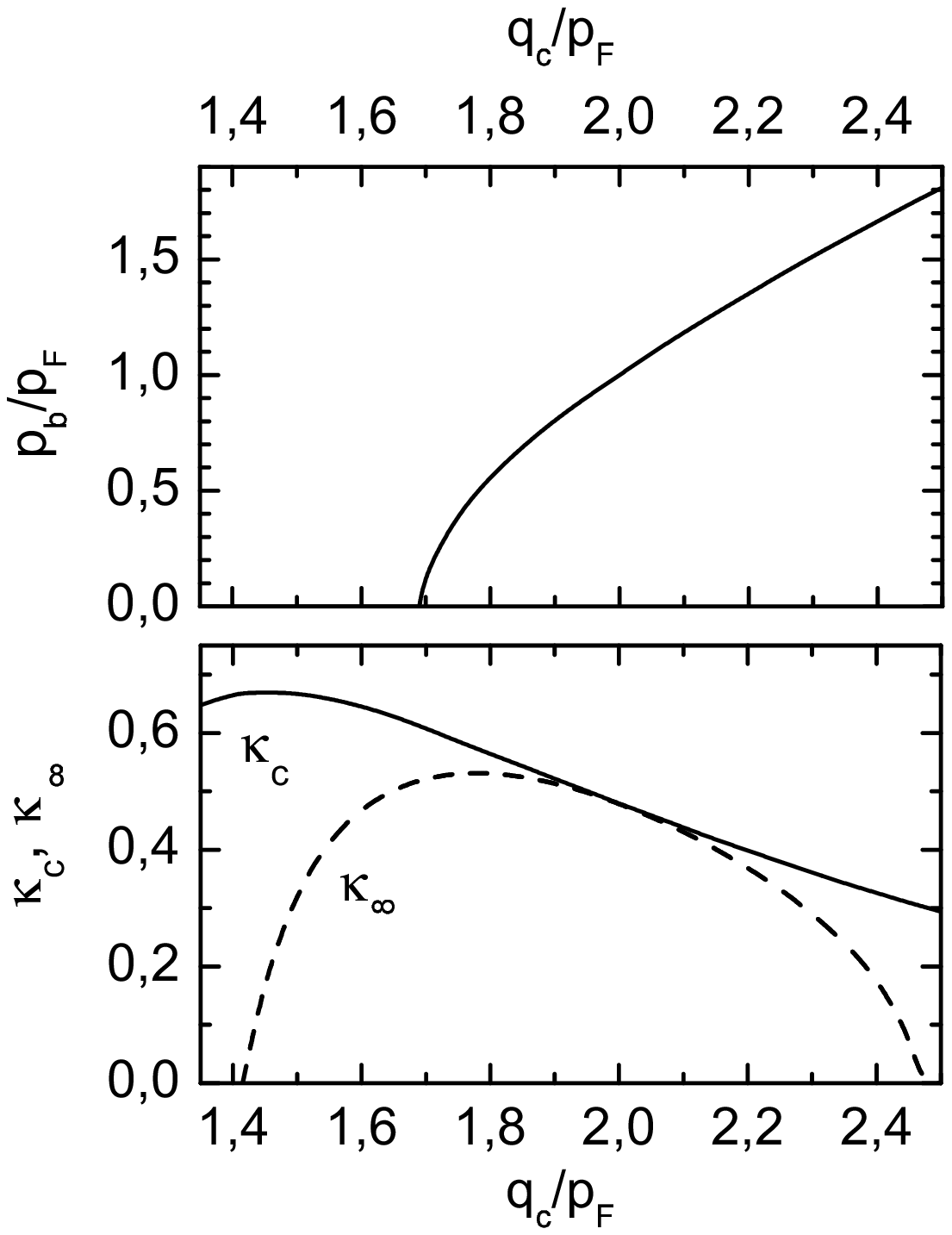,width=0.39\linewidth,height=0.52\linewidth}}
\end{center}
\caption{Same as in figure~\ref{fig:nmbe} but for 3D electron gas
}
\label{fig:3degbe}
\end{figure}

The spectrum $\xi(p)$ of the 3D electron gas, calculated for a critical
momentum $q_c=1.95\, p_F$ at three values of the parameter $\kappa$,
is drawn in panel (a) of figure~\ref{fig:3degxi}.  The solid line
shows the spectrum at $\kappa\simeq 0.5015$, for which equation~(\ref{root})
has a single root at the Fermi momentum $p_F$. The long-dashed line depicts
$\xi(p)$ at $\kappa=\kappa_b\simeq 0.5005$.  As seen, the bifurcation
point $p_b$ appears close to the Fermi momentum $p_F$.  Also shown
is the case when the bifurcation point reaches $p_F$: the short-dashed
line traces the sp spectrum at $\kappa_{\infty}\simeq 0.4987$, where the
effective mass becomes infinite.  This result was first obtained in
reference~\cite{ksz}.  The relevant plots for $q_c=2.0\, p_F$ are
displayed in panel (b). The solid line shows the spectrum at
$\kappa\simeq 0.4800>\kappa_b$.  For this choice of $q_c$, the
bifurcation point $p_b$ appears exactly at the Fermi surface when
$\kappa_b\simeq 0.4794$, as indicated by the long-dashed line.  Since
the effective mass goes to infinity, $\kappa_{\infty}$ and $\kappa_b$
coincide.  The short-dashed line corresponds to a case beyond the critical
point, with $\kappa\simeq 0.4780$.  In all three cases, in the spectrum
has a cubic-like shape as a function of $p-p_F$ in the vicinity of
the Fermi momentum.  The spectra for $q_c=2.05\,p_F$ are shown in
panel (c).  The solid line corresponds to $\kappa\simeq 0.4600$;
the long-dashed line, to $\kappa_b\simeq 0.4586$; and the dotted
line, to $\kappa_{\infty}\simeq 0.4568$.  The bifurcation point
is outside the Fermi sphere, with $p_b\sim 1.1\,p_F$.  The effective
mass goes to infinity at $\kappa_{\infty}<\kappa_b$, as in the case
of $q_c=1.95\,p_F$.  The behaviors of the parameters $p_b,\kappa_b$,
and $\kappa_{\infty}$ as functions of the critical momentum $q_c$ are
exhibited in figure~\ref{fig:3degbe}.  Analogous results are obtained
in a study of ferromagnetic fluctuations, when $q_c=0$.

\begin{figure}
\begin{center}
\mbox{\epsfig{file=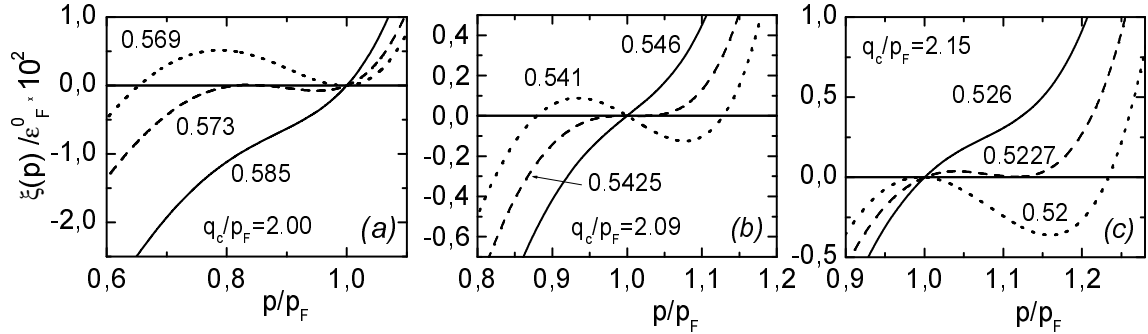,width=1.\linewidth,height=0.31\linewidth}}
\end{center}
\caption{
The same as in figure \ref{fig:3degxi} but for the 2D case.
}
\label{fig:2degxi}
\end{figure}

Calculation of the quasiparticle spectrum for 2D electron gas, with
results collected in figure~\ref{fig:2degxi}, confirms the implicit
judgement that the 2D problem does not differ qualitatively
from the 3D situation.

\section{Competition between different rearrangement scenarios}

We now turn the discussion to the proposed scenarios for alteration
of the Landau state beyond the limit of its stability, assuming that
the difference $|\rho-\rho_b|$ is much smaller than $\rho$.

\subsection{Ignoring the feedback effects of rearrangement:
Lifshitz-bubble formation}

In the pion-condensation example where $q_c\lsm p_F$, we have
seen that new roots of equation~(\ref{root}) arise quite far from
the Fermi momentum $p_F$.  As was shown in reference~\cite{vkzc},
the basic rearrangement mechanism transforming the Landau state
in this case involves the formation of some number of the Lifshitz
bubbles.  The quasiparticle occupation numbers $n(p)$ remain
integral at 0 or 1, but the Fermi surface becomes
multi-connected \cite{vary,zb}.

\begin{figure}
\begin{center}
\mbox{\epsfig{file=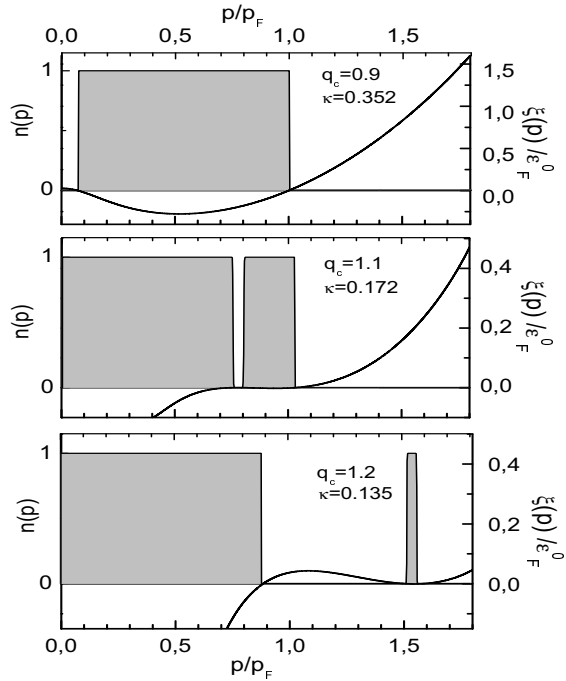,width=0.5\linewidth,height=0.6\linewidth}}
\end{center}
\caption{
Neutron spectra $\xi(p)$ (in units of $\varepsilon^0_F$),
together with the corresponding momentum distributions $n(p)$,
calculated for $q_c=0.9 p_F$ and $\kappa=0.352$ (panel (a)),
for $q_c=1.05 p_F$ and $\kappa=0.172$ (panel (b)), and
for $q_c=1.2 p_F$ and $\kappa=0.135$ (panel (c)).
}
\label{fig:nmbub}
\end{figure}

Figure~\ref{fig:nmbub} presents some results from calculations designed
to illustrate the characteristics of bubble formation.  The sp spectra
are evaluated by solving the closed equation for $\xi(p,T)$ that is
obtained upon substitution of the Fermi-Dirac distribution
$n(p,T)=[1+\exp(\xi(p,T)/T)]^{-1}$ into the r.h.s.\ of
equation~(\ref{pgr}). Tuning of the chemical potential $\mu$ is
governed by the normalization condition.  A very small temperature
($T=10^{-5}\varepsilon^0_F$) is used to imitate the zero-temperature
case.  The three panels in figure~\ref{fig:nmbub} show the neutron sp
spectra calculated at three values of the critical momentum of
spin-isospin fluctuations, $q_c=0.9\,p_F$ (panel (a)), $q_c=1.1\,p_F$
(panel (b)), and $q_c=1.2\,p_F$ (panel (c)). For all three parameter
choices, the density $\rho$ is slightly above the critical value, and
$n(p)$ exhibits a single LB, the position of which strongly depends
on $q_c$.  The bubble is located at the origin for $q_c=0.9\,p_F$,
at $p\sim 0.7\,p_F$ for $q_c=1.1\,p_F$, and mostly outside the original
Fermi sphere for $q_c=1.2\,p_F$.  The bubble is small in cases (a)
and (b), and the net disturbance relative to the original filled
Fermi sea is small in all three cases.

\begin{figure}
\begin{center}
\mbox{\epsfig{file=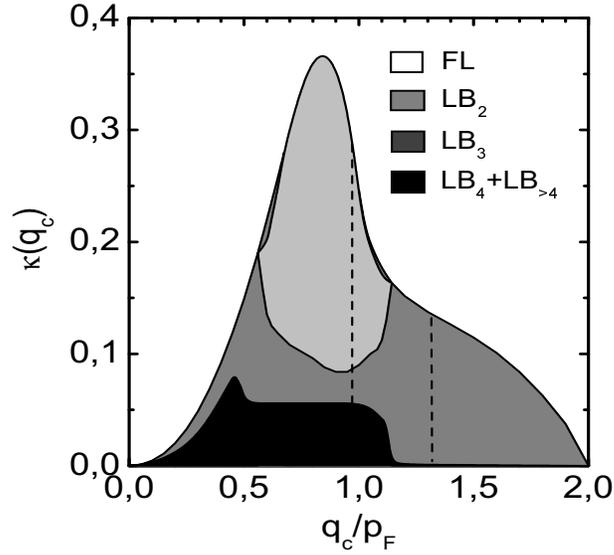,width=0.52\linewidth,height=0.48\linewidth}}
\end{center}
\caption{
The phase diagram of neutron matter in $(q_c,\kappa)$ variables.
The Landau phase (FL) occupies the white region of the plane. The
LB phases are denoted by LB$_i$, the index $i$ indicating the
number of sheets (or branches) of the Fermi surface.
Dashed lines show the borders of the territory occupied by the FC
provided the feedback is taken into account.
}
\label{fig:nmpha}
\end{figure}

As the density increases, the LB moves and multiplies. This
behavior is demonstrated in figure~\ref{fig:nmpha}, which shows
the phase diagram of neutron matter in the $(q_c,\kappa)$ plane.
The Landau state with $n(p)=n_F(p)$ occupies the white region of
the diagram (labeled FL in the figure).  The LB phases populate
the shaded part of the plane, which is separated from the FL
region by the curve $\kappa_b(q_c)$ (see figure~\ref{fig:nmbe}).
We shall not delve deeply into the ``zoology'' of the LB world,
instead classifying the LB phases simply by the number $i$ of
sheets of the Fermi surface.

Formation of Lifshitz bubbles is by no means the only kind
of rearrangement the Fermi surface can experience as a result
of the violation of stability condition (\ref{nsc}).  If the
bifurcation point in equation~(\ref{root}) is situated close to
the Fermi momentum $p_F$, then a new rearrangement scenario, fermion
condensation \cite{ks,vol,noz,physrep}, comes into play. Its salient
features are apparent from the basic equation
\beq
  {\delta E\over \delta  n(p)}=\mu , \qquad p\in {\rm C} \ .
\label{var}
\eeq
This equation determines a new quasiparticle distribution $n_0(p)$
that differs from the Fermi distribution $n_F(p)$ within the region C,
but coincides with it outside.  In contrast to the Lifshitz-bubble
phases, the rearranged distribution $n_0(p\in {\rm C})$ appears to
be a continuous function of $p$, with values lying {\it between} 0 and 1.
Since its l.h.s.\ is nothing but the quasiparticle energy $\epsilon(p)$,
the condition (\ref{var}) implies the presence of a completely flat
portion of the spectrum $\xi(p)$.  This plateau in $\xi(p)$ identifies
the fermion condensate (FC), i.e., the subsystem of quasiparticles
with energy pinned to the chemical potential.  As consequence of this
behavior, the density of states $\rho(\varepsilon)$ acquires an infinite
term at $\varepsilon=0$, as in a Bose liquid.  It must be kept in mind,
however, that the fermion condensation is in actuality an intermediate
stage, since its inherent degeneracy must somehow be lifted.
The analysis of this process is beyond the scope of the present article;
a detailed treatment may be found in reference~\cite{physrep}.

Equation (\ref{var}) can be rewritten in explicit form by employing
the well-known Landau formula
\beq
\delta E =\delta^{(1)} E+\delta^{(2)} E
\eeq
for the variation of the ground state energy $E$ under variation
$\delta n_F(p)=n(p)-n_F(p)$ of the Landau quasiparticle momentum
distribution $n_F(p)$. Here
\begin{eqnarray}
\delta^{(1)} E&=&
\int \xi(p;n_F)\,\delta n_F( p)\, d\tau\  \,,\nonumber \\
\delta^{(2)} E&=&{1\over 2}\,\int\!\!\int f({\bf p},{\bf p}_1;n_F)\,
  \delta n_F(p)\,\delta n_F(p_1)\,d\tau\,d\tau_1  \,,
\label{dele}
\end{eqnarray}
where $f$ is the Landau amplitude entering equation~(\ref{pgr}).
Insertion of this formula into condition (\ref{var}) leads to the
following equation for determining the new momentum distribution
$n_0(p)$,
\beq
\xi(p;n_F) + \int f({\bf p},{\bf p}_1;n_F)\left[n_0(p_1)-
  n_F (p_1)\right]d\tau_1=0 \ .
\label{varxi}
\eeq
Solutions of this equation can be assigned an order parameter
$\eta$, taken as the ratio of the FC density to the total density
$\rho$.  Nontrivial solutions can arise beyond the point where
the effective mass $M^*$ changes its sign.  However, as we
know from figure~\ref{fig:2degxi}, Lifshitz bubbles already exist
at this point.  Thus, in the model adopted, LB states make their
appearance prior to the formation of a fermion condensate.

To elucidate the situation, we may exploit the fact that in
the region adjacent to the Fermi momentum $p_F$, the group velocity
$d\xi/dp$ has essentially a parabolic shape.  Defining a new
variable $y= (p-p_F)/p_F$, we can write
\beq
{d\xi(y)\over dy}\simeq {p_F^2\over M}\,A\left(3(y-y_m)^2+b\right)\,,
\label{gr}
\eeq
\beq
\xi(y;n_F) =
{p^2_F\over M}\,A\left(
(y-y_m)^3+by+y_m^3\right)
\equiv
{p^2_F\over M}\,A y\left(y^2-3yy_m+3y_m^2+b\right) \ .
\label{sp}
\eeq
The three parameters $y_m$,
\beq
A={ M\over 6p^2_F}\,\left.{d^3\xi(y)\over dy^3}\right|_{p_F}>0\ ,
\quad
{\rm and} \quad b={M\over p_F^2 A}\,\left.{d\xi(y)\over dy}\right|_{y_m} <0
\label{coef}
\eeq
specifying the spectrum $\xi(p)$ depend on the parameter $\kappa$
appearing in the model form (\ref{tio}) for the Landau amplitude $f$.
We observe that the parameter $b$ must be negative in the vicinity
of the Fermi surface.  At the point $\kappa=\kappa_{\infty}$ where
the effective mass diverges, i.e. $(d\xi/dy)_F=0$, the parameters
$y_m$ and $b$ are connected by the relation
\beq
3y_m^2(\kappa_{\infty}) +b(\kappa_{\infty}) =0\, .
\label{bon}
\eeq
On the other hand, the equation $\xi(y)=0$, with $\xi(y)$ given by the
formula (\ref{sp}), has the single root $y=0$ for those $\kappa$
values at which
\beq
s_{LB}(\kappa)={3y_m^2(\kappa)\over 4}+b(\kappa) >0 \ .
\label{relb}
\eeq
Otherwise, the function $\xi(y)$ acquires two additional zeroes
\beq
y_{1,2}= {3y_m(\kappa)\over 2}\pm \left[-\left({3y_m^2(\kappa)\over 4}
  +b(\kappa)\right)\right]^{1/2} \,,
\label{rt}
\eeq
rendering the Landau state unstable. Setting $\kappa=\kappa_{\infty}$
in equation~(\ref{sp}) and appealing to relation (\ref{bon}), we infer
that at the point where fermion condensation sets in, the equation
$\xi(y)=0$ already has three roots, namely $y_{1,2}=0$ and
$y_3=3y_m(\kappa_{\infty})$.  This confirms that the Landau state
is unstable at the point of fermion condensation.

Thus, we have demonstrated both numerically and analytically that
in the over-simplified model under consideration, alteration of the
Landau state due to formation of Lifshitz bubbles does indeed
precede fermion condensation.  This property was first documented
in the numerical calculations of reference~\cite{zb}.

\subsection{A simple model including feedback: the contest between
fermion condensation and Lifshitz-bubble creation}

To this point, no consideration has been given to the effect of
feedback on the critical fluctuations as reflected in their
basic parameter, the stiffness coefficient $\kappa^2$ entering
the interaction function $f(q)$ of equation (\ref{tio}).  We now
address this issue.  Our analysis shows that the impact of
Lifshitz-bubble formation on the critical fluctuations is
insignificant.  On the other hand, the feedback effect may be
crucial in the case of fermion condensation, because of the
infinite value taken by the density of states $\rho(\varepsilon=0)$
at $T=0$.

To provide a basis for analysis, we evaluate the gain in energy due to
the emergence of a small FC fraction, assuming a trial FC function for
the variation $\delta n(y)=n_{\tri}(p)-n_F(p)$ having the simplest form,
\beq
\delta_{\tri} n(y)={1\over 2} \,{\rm sgn\,} y \ , \qquad
-{\lambda<y<\lambda} \ .
\label{tr}
\eeq
Particle number is conserved as long as the parameter
$\lambda$ is sufficiently small.  With this trial function, we
evaluate the first- and second-order variations,
$\delta^{(1)}_{\tri}E$ and $\delta^{(2)}_{\tri} E$, in
the Landau formula (\ref{dele}).

After inserting the trial function $\delta_{\tri} n(p)$ along with
the sp spectrum (\ref{sp}) into equation~(\ref{dele}), simple
manipulations yield
\beq
\delta^{(1)}_{\tri} E(\lambda)={p^2_F\over M}\,
{\lambda^2\over 4}\left[A\lambda^2+2v_g\right]     \  ,
\label{del1}
\eeq
and
\beq
\delta^{(2)}_{\tri} E={p^2_F\over M}\,{B\lambda^4\over 4} \ ,
\label{del2}
\eeq
where we have introduced the dimensionless group velocity
$v_g=A(3y^2_m+b)$.  Collecting terms, we arrive at
\begin{equation}
\delta_{\tri} E={p^2_F\over M}{\lambda^2\over 4}\left[(A+B)\lambda^2
+2v_g\right]   \ ,
\label{summ}
\end{equation}
with $v_g= s_{LB}+9y^2_m/4$ and $s_{LB}$ given by relation (\ref{relb}).

As we have seen, the LB phase wins the contest with the Landau state
if $s_{LB}<0$.  To uncover the conditions under which the FC state
can prevail in the competition between the two phases, let us
investigate the roots of the function $\delta_{\tri} E(\lambda)$
given by equation~(\ref{summ}).  Quite evidently, if $v_g>0$, or
equivalently, if $s_{LB}>-9y^2_m/4$, this function has no roots,
and hence $\delta_{\tri} E(\lambda)>0$.  This result demonstrates
that without accounting for feedback of the FC on the stiffness
coefficient $\kappa^2$, and hence on $v_g$, the FC phase
looses the contest.

To proceed further we make the simple assumption that $v_g$ falls
off linearly with increase of the FC density.  Thus we write
$v_g(\kappa,\lambda)=v_0(\kappa)-\lambda v_1(\kappa)$,
where $v_1(\kappa)>0$ is a slowly varying function of $\kappa$.
It is straightforward to show that equation $\delta_{\tri} E=0$
has two {\it positive} roots
\beq
\lambda_{1,2} =
{v_1\pm \sqrt{v_1^2-2(A+B)v_0}\over A+B}\ ,
\label{rt2}
\eeq
between which $\delta_{\tri} E(\lambda)<0$ holds provided
$v^2_1>2(A+B)v_0$.  Therefore for any $\lambda$
within the range $\lambda_1<\lambda<\lambda_2$, the variation
$\delta_{\tri} E(\lambda)$ is negative.  Since the true value
of $\delta_0 E$, calculated with the true function $n_0(p)$
from equation~(\ref{varxi}), should lie lower, we infer that fermion
condensation is energetically preferred over the Landau state
-- at least in the case that $v^2_1>2(A+B)v_0$.
This inequality is always satisfied close to the point of fermion
condensation, where according to equation~(\ref{bon}), $v_0$
vanishes.  Since both of the roots $\lambda_{1,2}$ are positive,
not zero, fermion condensation is predicted to be a weak first-order
phase transition.

In deciding the competition between the FC and LB phases, it is
instructive to focus on the case of small positive $s_{LB}(\kappa)$,
for which LB formation is still forbidden.  The input parameters
may be chosen so as to locate the minimum of $d\xi/dp$ not far from
the Fermi surface, which implies a sufficiently small value of $y_m$.
But at the point where $s_{LB}=0$, we have $v_0=9y^2_m/4$.
Hence, if $y_m$ is sufficiently small, both of the roots $\lambda_1$,
$\lambda_2$ of equation~(\ref{rt2}) are real, and
$\delta_{\tri} E(\lambda)$ is negative in the interval between
them.  We then conclude that for $s_{LB} \to 0^+$, fermion
condensation is allowed, while Lifshitz-bubble formation is
forbidden.

\subsection{ Numerical illustration }

\begin{figure}
\begin{center}
\mbox{\epsfig{file=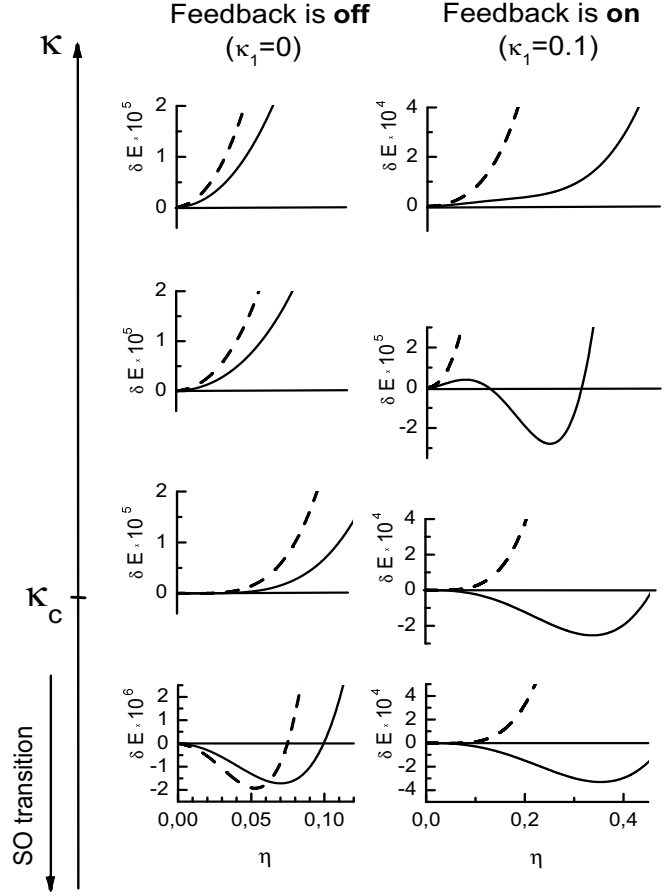,width=0.6\linewidth,height=0.8\linewidth}}
\end{center}
\caption{Energies (per particle) of the trial state $\delta_{\tri}E$
(solid line) and of the LB state $\delta_{\lb}E$ (dashed line),
calculated for the 3D electron gas with $q_c=2\,p_F$ and
$\kappa_0=0.4820,\, 0.4810, \, 0.4794$, and $0.4790$ (ordered
from top to bottom). The energies, measured in units of the
Fermi energy $\varepsilon^0_F$, are plotted versus the order
parameter$\eta$.  Left panels: feedback off ($\kappa_1=0$).
Right panels: feedback on ($\kappa_1=0.1$).
}
\label{fig:feed}
\end{figure}

The foregoing model analysis of the role of feedback in the
competition between fermion condensation and Lifshitz-bubble
formation can be illustrated by numerical calculation of the
variation $\delta E$ of the ground-state energy corresponding to
chosen variations $\delta n(p)$ of the quasiparticle distribution
away from the Fermi distribution $n_F(p)$.  (The same exercise
will serve to demonstrate that the parameter $\alpha$ appearing
in the second-order energy variation (\ref{del2}) is indeed of
order unity.  We compare the energy variation corresponding to
the FC trial variation $\delta_{\tri}n(p)$, with the energy variation
$\delta_{\lb}$ associated with the LB phase.  In figure~\ref{fig:feed},
the energy shifts $\delta_{\tri}E$ and $\delta_{\lb}E$, evaluated
for the 3D electron gas with $q_c=2\,p_F$, are drawn as functions
of the order parameter $\eta$, taken as the relative phase volume
of the region in momentum space within which the quasiparticle
distribution is rearranged.  The left panels show the results
obtained ignoring the suppression of the stiffness coefficient
$\kappa^2$ due to formation of the FC.  In this case, it is seen
that both the FC trial state and the LB state give lower energy
than the Landau state at $\kappa<\kappa_b$, but the LB state
has the deeper minimum.

The feedback of the quasiparticle rearrangement on the charge
fluctuations strongly alters the competitive balance between LB
and trial FC states.  For the trial FC state, the feedback effect
is included in the same manner as detailed above.  In particular,
we assert a linear dependence on $\eta$ of the term $\kappa^2$
in the denominator of the amplitude (\ref{tio}), in the form
$\kappa^2(\eta) = \kappa^2_0-\eta\,\kappa_1^2$.  To be definite,
we set $\kappa_1=0.1$.  For the LB state, feedback is unimportant,
since the density of states $\rho(\varepsilon)$ receives no dramatic
enhancement in this rearrangement scenario.  The right panels in
the figure demonstrate the role of feedback in the competition
between the three competing states.  We observe that the plots
of $\delta_{\tri}E(\eta)$ differ markedly from their counterparts
in the left panels, which represent the feedback-off situation.
In accordance with the analysis of the previous section, a
negative minimum of the curve $\delta_{\tri} E$ first appears
at a value of $\kappa$ below the critical value $\kappa_c$.  Beyond
$\kappa_c$ (right bottom panel), this minimum is lower than that of
the LB curve by two orders of magnitude. Therefore, the state
possessing the true FC -- whose energy is necessarily below that
of the trial FC state -- clearly wins the contest with the LB
phase, and the transition from the Landau state to the FC state
is of first order.

Taking the feedback into account changes the phase diagram of
neutron matter. The FC wins the contest with the LB states in
the part of non-Landau area of the $(q_c,\kappa)$
phase diagram between two dashed lines in figure \ref{fig:nmpha}.
We estimated these borders using the same parameter $\kappa_1=0.1$
as for the 3D electron gas.

\section{Conclusions}

Based on standard relations of the Landau theory of Fermi liquids,
we have explored the properties of mechanisms that may force a
rearrangement of the Fermi surface of a homogeneous system at zero
temperature.  It is found that in advance of a second-order phase
transition to a state with long-range order induced by the softening
of the spectrum of critical fluctuations, there arise additional,
nontrivial roots of the equation $\epsilon(p)=\mu$, signaling an
instability of the Landau state.  The consequent metamorphosis of
the quasiparticle spectrum has been traced to a divergence, at the
second-order transition point, of the leading term in the quasiparticle
amplitude,  which is proportional to the pertinent static
susceptibility.

We have clarified the competitive status of two scenarios for
alteration of the Landau state, Lifshitz-bubble formation and
fermion condensation.  In general, and in particular for the
case of fermion condensation, it must be expected that the
rearrangement of the quasiparticle momentum distribution will
exert an influence on the implicated collective degree of freedom.
This feedback effect has been taken into account through a
simple model in which the stiffness coefficient
depends linearly on the FC density.  Without feedback,
Lifshitz-bubble formation precedes fermion condensation.
However, the introduction of feedback reverses this picture:
under increase of density, the first-order phase transition
to the state containing a fermion condensate takes place
before bubble-formation becomes the favored state.

\subsection{Acknowledgments}

This research was supported in part by the National Science Foundation
under Grant No.~PHY-0140316 (JWC and VAK), by the McDonnell Center
for the Space Sciences at Washington University (VAK), by Grant
No.~NS-1885.2003.2 from the Russian Ministry of Industry and
Science (VVB, VAK, and MVZ), by Fellow Grants from Russian Research
Centre ``Kurchatov Institute'' (VVB and MVZ) and by a Student
Grant from the ``Dynasty'' Foundation (VVB).  Two of the authors
(VVB and MVZ) thank INFN (Sezione di Catania) for hospitality
during their stay in Catania.


\section*{References}
\begin{thebibliography} {99}

\bibitem{lan} Landau L D 1956 {\it Sov. Phys.--JETP} {\bf 30} 1058 \\
              Landau LD 1958  {\it Sov. Phys.--JETP} {\bf 35} 97

\bibitem{mig0} Migdal A B 1967 {\it Theory of Finite Fermi Systems and
Applications to Atomic Nuclei} (New York: Wiley)

\bibitem{cas} Casey A, Patel H, Nyeki J, Cowan B P and Saunders J
 2003 \PRL {\bf 90} 115301

\bibitem{krav}  Kravchenko S V and Sarachik M P 2004
  Rep. Prog. Phys. {\bf 67} 1

\bibitem{pud} O. Prus, Y. Yaish, M. Reznikov, U. Srivan and V. M. Pudalov
2003 \PR {\bf B\, 67} 205407

\bibitem{vary} De Llano M and Vary J P (1979) \PR {\bf C\,19} 1083 \\
  De Llano M, Plastino A and Zabolitsky J G (1979) \PR {\bf C\,20} 2418

\bibitem{aguil} Aguilera-Navarro V C, De Llano M, Clark J W and
Plastino A (1982) \PR {\bf C\,25} 560

\bibitem{ks} Khodel V A and Shaginyan V R (1990) {\it JETP Lett.} {\bf 51}
553

\bibitem{vol} Volovik G E (1991) {\it JETP Lett.} {\bf 53} 222
	      Volovik G E (1994) {\it JETP Lett.} {\bf 59} 830

\bibitem{noz} Nozieres P (1992) \JP I {\bf 2} 449

\bibitem{physrep} Khodel V A, Shaginyan V R and Khodel V V (1994)
{\it Phys. Rep.} {\bf 249} 1

\bibitem{ksz} Khodel V A, Shaginyan V R and Zverev M V (1997)
{\it JETP Lett.} {\bf 65} 254

\bibitem{mig1} Migdal A B (1978) \RMP {\bf 50} 107

\bibitem{mig2} Migdal A B, Saperstein E E, Troitsky M A and
Voskresensky D N (1998) {\it Phys. Rep.} {\bf 192} 179

\bibitem{pand} Akmal A, Pandharipande V R and Ravenhall D G
(1998) \PR {\bf C\,58} 1804

\bibitem{wir} Wiringa R B, Fiks V and Fabrocini A (1988)
{\it Phys. Rev.} {\bf 38} 1010

\bibitem{krot} Boronat J, Casulleras J, Grau V, Krotscheck E
and Springer J   (2003) Phys. Rev. Lett. {\bf 91} 085302

\bibitem{vkzc} Voskresensky D N, Khodel V A, Zverev M V and
Clark J W (2000) {\it Ap. J. Lett.} {\bf 533} 127

\bibitem{trio} Abrikosov A A, Gor'kov L P, Dzyaloshinskii I Y
1990 {\it Quantum Field Theoretical Methods in Statistical Physics}
(New York: Pergamon Press)

\bibitem{zb} Zverev M V and Baldo M 1998 {\it JETP} {\bf 87} 1129 \\
             Zverev M V and Baldo M 1999 \JPCM {\bf 11} 2059



\end{thebibliography}
\end{document}